\documentclass[lettersize,journal]{IEEEtran}
\usepackage{amsmath,amsfonts}
\usepackage{algorithmic}
\usepackage{algorithm}
\usepackage{array}
\usepackage[caption=false,font=normalsize,labelfont=sf,textfont=sf]{subfig}
\usepackage{textcomp}
\usepackage{stfloats}
\usepackage{url}
\usepackage{verbatim}
\usepackage{graphicx}
\usepackage{cite}
\usepackage{booktabs}
\usepackage{multirow}
\begin{document}

\title{CAT-MoEformer: Context-Aware Temporal MoE Transformer for Beam Prediction}

\author{Changkai Zhou, Cunhua Pan,~\IEEEmembership{Senior Member,~IEEE}, Hong Ren,~\IEEEmembership{Member,~IEEE}, Jiangzhou Wang,~\IEEEmembership{Fellow,~IEEE}

\thanks{Changkai Zhou, Cunhua Pan, Hong Ren and Jiangzhou Wang are with National Mobile Communications Research Laboratory, Southeast University, Nanjing 211189, China. (e-mail: 220251242, cpan, hren, j.z.wang@seu.edu.cn).}}

\maketitle

\begin{abstract}
This paper proposes CAT-MoEformer, a context-aware transformer with scene-conditioned mixture-of-experts (MoE) feed-forward networks, for proactive mmWave beam prediction from compressed uplink pilot observations. The spatial encoder comprises a three-layer asymmetric convolutional network followed by a squeeze-and-excitation recalibration block, which extracts frequency-beam correlation features from pilot tensors without explicit channel reconstruction. A truncated pretrained GPT-2 backbone models the temporal evolution of beam sequences, with the feed-forward networks in the upper three transformer layers replaced by scene-conditioned MoE-FFN modules. A lightweight gating network maps the scenario label and normalized user equipment speed to expert mixing weights, conditioning the routing decision on physical propagation descriptors rather than on latent hidden states. This design yields interpretable expert assignments and eliminates the load imbalance associated with token-level routing. To prevent expert collapse under soft routing, a three-stage training strategy is introduced: hard expert assignment in the first stage establishes scene-specific specialization, isolated gating network training in the second stage aligns the soft routing distribution with the hard partition, and top-1 hard inference in the third stage fine-tunes the model under deterministic single-expert activation to maximize scene-specific precision. Simulation results on 3GPP TR 38.901 Urban Macro channel simulations with $64{,}000$ user samples demonstrate that CAT-MoEformer achieves a Top-1 beam prediction accuracy of $94.88\%$ and a beam switching instant accuracy of $80.62\%$, representing gains of $2.33\%$ and $9.55\%$ respectively over a CNN+GPT-2 baseline, with an inference latency of $0.52$~ms.
\end{abstract}

\begin{IEEEkeywords}
beam prediction, mixture-of-experts, Millimeter-wave, scene-aware routing.
\end{IEEEkeywords}

\section{Introduction}

Millimeter-wave (mmWave) communications have emerged as a foundational technology for fifth-generation (5G) and beyond wireless networks, offering abundant spectrum to support multi-gigabit-per-second data rates~\cite{busari2018millimeter}. Signals in the mmWave band, however, suffer from severe free-space path loss and are highly susceptible to blockage~\cite{giordani2019tutorial}. To compensate for these propagation impairments, massive multiple-input multiple-output (MIMO) systems with highly directional beamforming are widely deployed~\cite{xue2024survey}. To establish and maintain robust directional links, the 3GPP New Radio (NR) standard specified a beam management framework comprising beam sweeping, measurement, determination, and reporting~\cite{giordani2019tutorial}. The use of narrow beams, however, introduces substantial pilot overhead during beam sweeping, which increases alignment latency and complicates beam switching. This problem is especially pronounced in high-mobility scenarios, where frequent handovers and radio link failures demand rapid and continuous beam tracking~\cite{xue2024survey, giordani2019tutorial, wang2024beam}.

Conventional beam alignment and tracking rely on codebook-based 
sweeping and classical signal processing. Early 3GPP NR beam 
management frameworks mainly adopted exhaustive or hierarchical beam 
sweeping strategies for initial access and beam tracking in mmWave 
systems~\cite{giordani2018initial, giordani2019standalone}. Exhaustive 
search methods can identify the optimal beam pair, but the associated 
overhead scales linearly with the antenna count, rendering them 
impractical for large-scale MIMO systems~\cite{lin2022novel}. To 
reduce alignment latency, model-based tracking algorithms such as the 
Kalman filter (KF) and extended Kalman filter (EKF) have been widely 
studied to exploit the temporal correlation of mmWave 
channels~\cite{jayaprakasam2017robust, chen2023mmwave}. Compressed 
sensing (CS) techniques have also been integrated into KF-based 
frameworks to address the grid mismatch problem and improve tracking 
accuracy under reduced pilot budgets~\cite{lin2022novel, 
shen2023compressed}, leveraging the angular-domain sparsity inherent 
in mmWave channels. Some schemes further incorporated side information 
such as UE position and spatial context to support proactive beam 
alignment under high mobility~\cite{chen2023mmwave}. These methods, 
however, rely on specific stochastic assumptions or rigid motion 
models, and their accuracy degrades substantially in non-stationary 
environments characterized by sudden blockages and irregular movement.

The limitations of model-based approaches have motivated a shift toward deep learning (DL) for intelligent beam management, owing to its capacity to capture complex non-linear mappings and non-stationary channel dynamics~\cite{khan2023machine}. Data-driven methods based on recurrent neural networks (RNNs) and long short-term memory (LSTM) networks exploit the temporal correlation of wireless channels for proactive beam prediction~\cite{zhang2024tbp, liu2025sa}. The integration of multi-modal side information has also been explored to improve robustness, including out-of-band spatial features from sub-6~GHz channels~\cite{ma2021deep} and sensory data such as GPS coordinates and camera images~\cite{dissanayake2023towards}. More recent work introduced flexible architectures that balance computational cost and accuracy across diverse propagation conditions, including multi-stage and speed-adaptive prediction networks~\cite{wang2025deep, wang2023deep}, as well as hierarchical beam alignment frameworks that learn coarse-to-fine probing codebooks to reduce measurement overhead while improving alignment accuracy~\cite{yang2024hierarchical}. By transitioning from reactive tracking to predictive inference, these frameworks reduce pilot overhead while maintaining link reliability in dynamic propagation environments. Despite these advances, two structural limitations persist. First, recurrent architectures update their hidden state sequentially, causing information from earlier frames to attenuate over long observation windows. This is particularly consequential at beam switching instants, where accurate prediction requires retaining long-range temporal context that LSTM-type models do not preserve reliably. Second, existing methods train a single model across all propagation conditions without distinction. LOS and NLOS channels differ substantially in angular spread, coherence time, and multipath structure, and allocating a shared parameter set uniformly across these heterogeneous regimes limits the degree of specialization achievable within each individual condition.

The advent of large pretrained transformer models has opened a direct avenue for addressing these limitations. Recent studies demonstrated that transfer learning with pretrained transformer architectures can effectively capture complex wireless channel dynamics and generalize across heterogeneous propagation environments~\cite{Hu2024transfer}. Recent work further formulated beam prediction as a time-series forecasting task and showed that pretrained LLM backbones outperform task-specific recurrent models in both accuracy and robustness~\cite{sheng2025beam}. For near-field beam tracking in extremely large-scale antenna systems, a CNN-GPT-2 framework was proposed to model abrupt, non-linear temporal beam dynamics~\cite{liu2025large}. 

To handle heterogeneous multi-modal inputs across multiple frequency bands, MoE architectures with adaptive gating have been applied to decouple feature extraction across modalities and reduce negative transfer~\cite{zhang2025multimodal}. The modern sparse MoE paradigm was largely established by the sparsely-gated MoE layer proposed in~\cite{shazeer2017outrageously}, which introduced conditional computation and top-$k$ expert routing for scalable transformer modeling. Building upon this idea, parameter-efficient MoE adaptation has further been explored for wireless communication systems and beam prediction tasks~\cite{lei2026llm, liu2025llm4wm}. These works collectively established the feasibility of combining pretrained transformers with expert routing for beam management tasks. 
A gap nonetheless remains. Existing MoE formulations either apply token-level dynamic routing derived from hidden states, which is susceptible to expert collapse and load imbalance in small-scale systems, or route without conditioning on explicit physical propagation descriptors. Neither approach exploits the structured heterogeneity of the channel environment—specifically, the distinct temporal dynamics arising from LOS versus NLOS propagation and from differences in UE mobility—as a direct routing signal. To the best of the authors' knowledge, no prior work jointly addresses proactive beam prediction from compressed pilot observations, pretrained transformer-based temporal modeling, and physically grounded scene-adaptive expert routing within a single framework.

This paper proposes CAT-MoEformer, a Context-Aware Transformer with scene-conditioned Mixture-of-Experts feed-forward networks (FFN), for proactive mmWave beam prediction. The main contributions of this paper are summarized as follows:

\textbf{Physically-Grounded Scene-Conditioned MoE Architecture:} We propose CAT-MoEformer, a novel framework that leverages a truncated pre-trained GPT-2 backbone for beam sequence modeling. By replacing the conventional FFN in the upper transformer layers with scene-conditioned Mixture-of-Experts (MoE-FFN) modules, the architecture effectively captures spatial-temporal channel dynamics. Unlike token-level routing, our lightweight gating network utilizes physical propagation descriptors—specifically scenario labels and normalized UE speeds—as inputs. This design ensures interpretable expert assignments and inherently eliminates load imbalance and the need for auxiliary loss functions.

\textbf{Staged Expert Specialization and Routing Optimization Strategy:} To mitigate the pervasive issue of expert collapse in MoE training, we develop a three-stage hierarchical training strategy. The first stage enforces hard expert assignments based on ground-truth scene labels to establish independent specialization for each expert. In the second stage, the gating network is trained to align the soft routing distribution with the previously established hard partitions. Finally, targeted fine-tuning of the expert output layers and the classifier is performed to achieve seamless domain adaptation and maximize predictive precision.

Extensive simulation results on 3GPP TR 38.901 Urban Macro channel simulations demonstrate that CAT-MoEformer achieves substantial gains in both overall Top-1 accuracy and beam switching instant accuracy over CNN, CNN+LSTM, and CNN+GPT-2 baselines, while maintaining an inference latency well within the slot-level coherence budget of 3GPP NR.

The remainder of this paper is organized as follows.
Section~\ref{sec:system} presents the system model and problem formulation. Section~\ref{sec:model} describes the proposed CAT-MoEformer architecture and three-stage training strategy.
Section~\ref{sec:simulation} presents simulation results, ablation studies, and architecture design analysis.
Section~\ref{sec:conclusion} concludes the paper.

\section{System Model}
\label{sec:system}
\subsection{Scenario and Channel Setup}

Consider an uplink mmWave communication system operating at a carrier frequency of $f_c = 28$~GHz, which falls within the 3GPP FR2 spectrum. The base station (BS) is equipped with an $N_r = 64$-element uniform linear array (ULA). The inter-element spacing is set to $d = \lambda/2$, where $\lambda = c/f_c$ denotes the carrier wavelength. Each user equipment (UE) is assumed to have a single omnidirectional antenna. The system employs an OFDM waveform~\cite{zhu2010adaptive, larsson2014mimo} with $K = 60$ active subcarriers and a subcarrier spacing of $\Delta f = 120$~kHz, yielding an effective bandwidth of $B \approx 7.08$~MHz.

UEs are randomly generated within a horizontal angular range of $\phi \in [-60^{\circ}, 60^{\circ}]$ relative to the array broadside and a radial distance range of $d_\text{UE} \in [30, 100]$~m from the BS. The mobility speed of each UE is drawn independently from a uniform distribution:
\begin{equation}
    v \sim \mathcal{U}(0,\ 120/3.6\ \text{m/s}).
    \label{eq:speed}
\end{equation}
Channels are generated in accordance with the 3GPP TR 38.901 Urban Macro (UMa) model under both line-of-sight (LOS) and non-line-of-sight (NLOS) conditions, distinguished by a binary scenario label $s \in \{0, 1\}$.

The channel is sampled at a uniform interval of $\Delta t = 10$~ms. Each UE produces $T+1 = 11$ consecutive snapshots, forming the channel tensor $\mathbf{H}^{(u)} \in \mathbb{C}^{N_r \times K \times (T+1)}$. The first $T = 10$ snapshots constitute the model input, while the $(T\!+\!1)$-th snapshot serves as the prediction target.

\subsection{Uplink Pilot Transmission}

To extract beam-relevant spatial features without relying on explicit CSI feedback, a compressed uplink sounding scheme based on hybrid precoding is adopted. The BS is equipped with $N_{rf} = 8$ RF chains.

\subsubsection{Wide-Beam Codebook}

A DFT-based wide-beam codebook comprising $S_w = N_r / m = 32$ codewords is constructed, where $m = 2$ adjacent DFT basis vectors are coherently combined per codeword. Let
\begin{equation}
    \mathbf{F}_{DFT} = \frac{1}{\sqrt{N_r}}\bigl[\mathbf{a}(\theta_0),\
    \mathbf{a}(\theta_1),\ \ldots,\ \mathbf{a}(\theta_{N_r-1})\bigr]
    \in \mathbb{C}^{N_r \times N_r}
    \label{eq:dft}
\end{equation}
denote the normalized DFT codebook. The $g$-th wide-beam codeword is defined as
\begin{equation}
    \mathbf{f}_{w,g} =
    \frac{\displaystyle\sum_{i=(g-1)m+1}^{gm}\mathbf{f}_{dft,i}}
    {\left\|\displaystyle\sum_{i=(g-1)m+1}^{gm}\mathbf{f}_{dft,i}\right\|},
    \quad g = 1,\ldots,S_w.
    \label{eq:widebeam}
\end{equation}
Each codeword covers $m$ adjacent spatial directions while satisfying the constant-modulus constraint of analog phase shifters. The full codebook $\mathbf{F}_{RF} \in \mathbb{C}^{N_r \times S_w}$ is applied in blocks of $N_{rf}$ columns over $n_{ofdm} = S_w / N_{rf}$ consecutive OFDM symbols per subframe.

\subsubsection{Digital Precoder and Pilot Design}

The digital precoder $\mathbf{F}_{BB,k} \in \mathbb{C}^{N_{rf} \times N_s}$ at subcarrier $k$ is constructed from a DFT unitary matrix with column permutations varying across subcarriers, providing frequency diversity. Zadoff-Chu (ZC) sequences with distinct root indices are employed as uplink pilots $\{x_{k,l}\}$, ensuring orthogonality across OFDM symbols.

\subsubsection{Received Signal Model}

At slot $t$, subcarrier $k$, and OFDM symbol $l$, the received signal after hybrid combining is
\begin{equation}
    \mathbf{y}_{t,k,l} = \sqrt{\frac{p_u}{K}}\,
    \mathbf{F}_{BB,k}^H \mathbf{F}_{RF,l}^H \mathbf{h}_{t,k}\, x_{k,l}
    + \mathbf{F}_{BB,k}^H \mathbf{F}_{RF,l}^H \mathbf{h}_{t,k}\, n_{t,k},
    \label{eq:received}
\end{equation}
where $\mathbf{h}_{t,k} \in \mathbb{C}^{N_r}$ is the uplink channel vector, $p_u$ is the total UE transmit power distributed equally across subcarriers, and $n_{t,k} \sim \mathcal{CN}(0, \sigma_n^2)$ is additive white Gaussian noise. Concatenating $\mathbf{y}_{t,k,l}$ over all $n_{ofdm}$ symbols yields the pilot feature vector $\mathbf{p}_{t,k} \in \mathbb{C}^{S_w}$. The per-slot observation matrix is
\begin{equation}
    \mathbf{P}_t = [\mathbf{p}_{t,1},\ldots,\mathbf{p}_{t,K}]^T
    \in \mathbb{C}^{K \times S_w}.
    \label{eq:Pt}
\end{equation}

\subsection{Beam Label Definition}

The optimal beam index at slot $t$ is the codeword maximizing the average received power across subcarriers:
\begin{equation}
    b_t^* = \underset{g \in \{1,\ldots,S_w\}}{\arg\max}\ \frac{1}{K}
    \sum_{k=1}^{K}\left|\mathbf{f}_{w,g}^H\,\mathbf{h}_{t,k}\right|^2.
    \label{eq:label}
\end{equation}
The prediction target is $b_{T+1}^*$, the optimal beam index of the subsequent slot. This one-step-ahead formulation targets the proactive handover scenario, in which beam selection must be committed before the channel transition occurs.
Beam indices with fewer than 700 samples are excluded for statistical reliability. The remaining $C$ valid classes are remapped to
$\{0, 1, \ldots, C-1\}$.

\subsection{Input Feature Representation}

The complex observation $\mathbf{P}_t$ is decomposed into real and imaginary parts and concatenated along a channel axis:
\begin{equation}
    \tilde{\mathbf{P}}_t = \bigl[\operatorname{Re}(\mathbf{P}_t)\ \big\|\
    \operatorname{Im}(\mathbf{P}_t)\bigr] \in \mathbb{R}^{2 \times K \times S_w}.
    \label{eq:split}
\end{equation}
Per-sample zero-mean unit-variance normalization is applied to suppress the influence of large-scale path loss. The complete input tensor for one UE is
\begin{equation}
    \mathbf{X} = \bigl(\tilde{\mathbf{P}}_1,\ \tilde{\mathbf{P}}_2,\ \ldots,\
    \tilde{\mathbf{P}}_T\bigr) \in \mathbb{R}^{T \times 2 \times K \times S_w},
    \label{eq:input}
\end{equation}
where $T=10$, $K=60$, and $S_w=32$. The scenario label $s$ and normalized speed $\bar{v} = 3.6v/120 \in [0,1]$ are appended as auxiliary inputs, broadcast along the time axis to match sequence length $T$.

It is worth emphasizing that $s$ and $\bar{v}$ are not redundant side information. The pilot tensor $\mathbf{X}$ conflates LOS/NLOS propagation characteristics and speed-induced Doppler variations into a single compressed observation, making these factors difficult to separate through a generic feature extractor alone. Supplying them explicitly provides the conditioning signal required by the context-aware routing mechanism.

\subsection{Problem Formulation}

Given the observation sequence $\mathbf{X}$ and context $(s, \bar{v})$, the beam prediction task is formulated as a multi-class classification problem:
\begin{equation}
    \hat{b}^* = \mathcal{F}_\theta\!\left(\mathbf{X},\ s,\ \bar{v}\right),
    \quad \hat{b}^* \in \{0,\ldots,C-1\},
    \label{eq:task}
\end{equation}
where $\mathcal{F}_\theta$ denotes the proposed model. Performance is evaluated by Top-1 accuracy $\mathbb{P}(\hat{b}^* = b_{T+1}^*)$ and Top-3 accuracy $\mathbb{P}(b_{T+1}^* \in \hat{\mathcal{B}}_3)$, where $\hat{\mathcal{B}}_3$ is the set of three highest-ranked beam candidates.

A particularly demanding evaluation subset consists of the \emph{beam switching instants}, defined by $b_T^* \neq b_{T+1}^*$. At such moments the channel undergoes an abrupt spatial transition, and accurate prediction directly reduces handover latency and improves link continuity. Accuracy on this subset is reported separately throughout the experimental evaluation.

\section{CAT-MoEformer Scheme}
\label{sec:model}
\subsection{Overall Architecture}

\begin{figure*}[t]
    \centering
    \includegraphics[width=\textwidth, trim=1bp 1bp 1bp 1bp, clip]{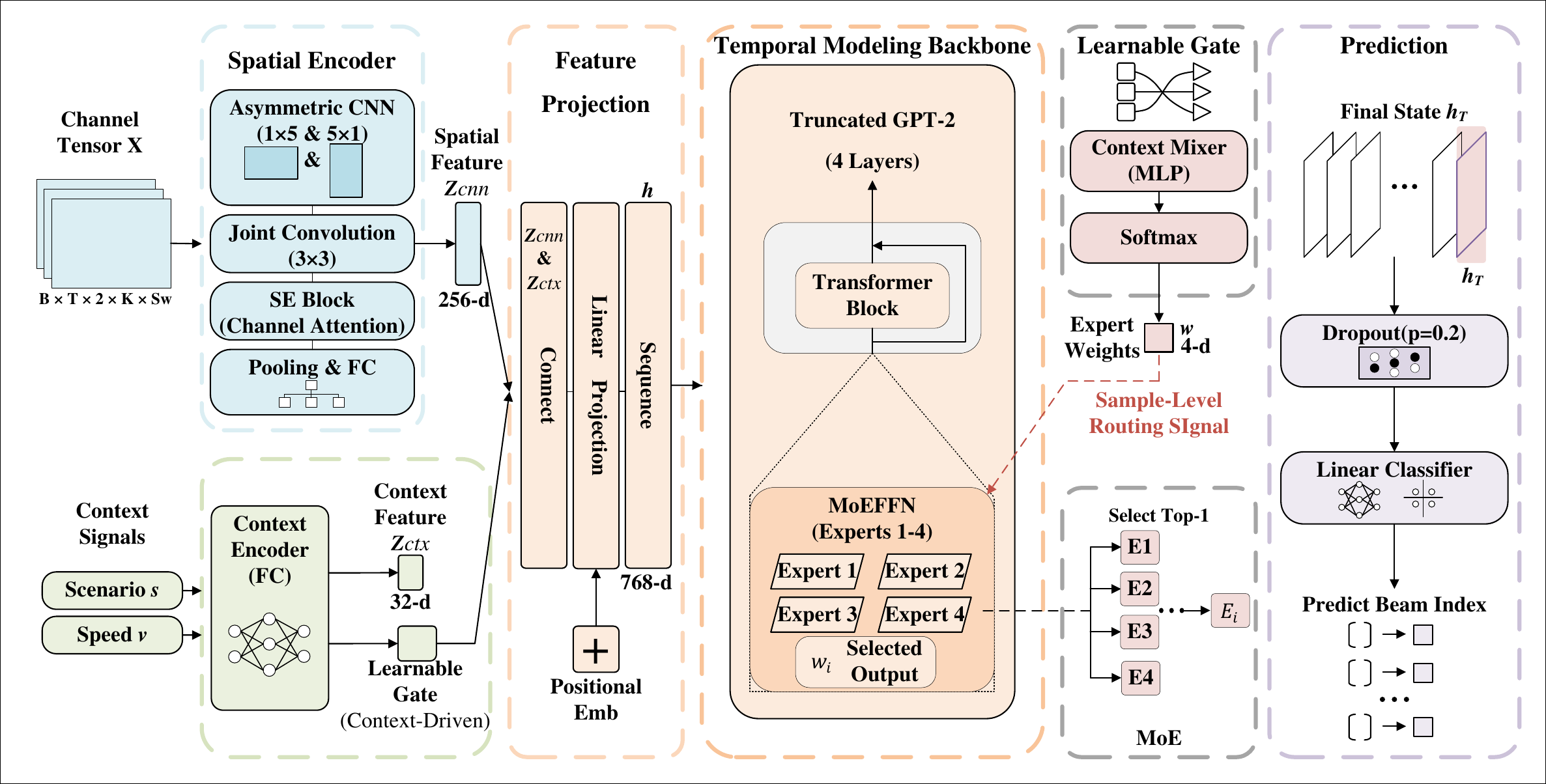}
    \caption{The overall architecture of our proposed CAT-MoEformer, featuring a spatial encoder, context fusion, and context-driven MoE temporal modeling.}
    \label{fig:architecture}
\end{figure*}

The proposed model, termed CAT-MoEformer, takes two types of input. The primary input is the pilot observation tensor $\mathbf{X} \in \mathbb{R}^{B \times T \times 2 \times K \times S_w}$, where $B$ denotes the batch size, $T = 10$ the number of observed slots, and the remaining dimensions correspond to the two-channel real-imaginary representation of the per-slot observation matrix defined in~\eqref{eq:split}. The auxiliary inputs are the scenario label sequence $\mathbf{s} \in \mathbb{R}^{B \times T}$ and the normalized speed sequence $\bar{\mathbf{v}} \in \mathbb{R}^{B \times T}$, both broadcast uniformly along the time axis so that every frame carries the same environmental context.

The forward pass consists of four sequential stages, as illustrated in Fig.~\ref{fig:architecture}. 
In the first stage, the time dimension of $\mathbf{X}$ is unfolded so that all $B \times T$ frames are processed independently by the spatial encoder. Each frame $\tilde{\mathbf{P}}_t \in \mathbb{R}^{2 \times K \times S_w}$ is passed through a three-layer asymmetric CNN followed by a squeeze-and-excitation (SE) attention block, producing a 256-dimensional spatial feature vector $\mathbf{z}_t^{\text{cnn}}$. The frame axis is then restored, yielding a sequence $\mathbf{Z}^{\text{cnn}} \in \mathbb{R}^{B \times T \times 256}$.
In the second stage, the context encoder projects the per-frame pair $(s_t, \bar{v}_t)$ into a 32-dimensional embedding $\mathbf{z}_t^{\text{ctx}}$. The spatial features and context embeddings are concatenated along the feature dimension and projected to the GPT-2 hidden dimension $d = 768$ via a single linear layer:
\begin{equation}
    \mathbf{e}_t = \mathbf{W}_p \bigl[\mathbf{z}_t^{\text{cnn}} \;\|\;
    \mathbf{z}_t^{\text{ctx}}\bigr] \in \mathbb{R}^{768}.
    \label{eq:proj}
\end{equation}
Learnable position embeddings from the pretrained GPT-2 vocabulary are then added to $\mathbf{e}_t$ to form the sequence input $\mathbf{h} \in \mathbb{R}^{B \times T \times 768}$.
In the third stage, the gating network reads the final-frame context $(s_T, \bar{v}_T)$ and produces a set of four expert mixing weights $\mathbf{w} \in \mathbb{R}^{B \times 4}$ via a two-layer MLP with softmax output. The sequence $\mathbf{h}$ is then passed through four transformer blocks. The first block retains the standard GPT-2 FFN for low-level feature transformation. The remaining three blocks replace their FFN with MoE-FFN layers, each receiving $\mathbf{w}$ as the routing signal. Within each MoE-FFN, the expert with the highest gating weight is selected, and the output of that expert alone constitutes the feed-forward output for the current sample:
\begin{equation}
    \text{MoE-FFN}(\mathbf{h}_t, \mathbf{w}) =
    E_{i^*}(\mathbf{h}_t), \quad
    i^* = \arg\max_{i \in \{1,2,3,4\}} w_i.
    \label{eq:moeffn_top1}
\end{equation} 
This top-1 hard selection is applied exclusively during Stage~3 fine-tuning and at inference time. It eliminates inter-expert interference in the feed-forward output and reduces the per-layer computation to that of a single expert network, yielding the lowest inference latency among the three routing modes employed across the training pipeline.
In the fourth stage, the hidden state of the final time step $\mathbf{h}_T \in \mathbb{R}^{B \times 768}$ is extracted and passed through a dropout layer with rate $p = 0.2$, followed by a linear classifier that produces $C$-dimensional logits. The predicted beam index is
\begin{equation}
    \hat{b}^* = \arg\max_{c} \; \mathbf{W}_{\text{cls}}\,
    \text{Dropout}(\mathbf{h}_T),
    \label{eq:cls}
\end{equation}
where $\mathbf{W}_{\text{cls}} \in \mathbb{R}^{C \times 768}$ is the classifier weight matrix. The final hidden state is used rather than a pooled representation because the causal self-attention in GPT-2 ensures that $\mathbf{h}_T$ already aggregates information from all preceding frames.

The overall computation can be written compactly as
\begin{multline}
    \hat{b}^* = \mathcal{F}_\theta\bigl(\mathbf{X},\, \mathbf{s},\,
    \bar{\mathbf{v}}\bigr) = \text{Cls}\!\left( \text{GPT2-MoE}\!\left(
    \text{Proj}\!\left([\mathbf{Z}^{\text{cnn}} \|\,
    \mathbf{Z}^{\text{ctx}}]\right) ,\right.\right. \\
    \left.\left. \text{Gate}(s_T, \bar{v}_T) \right) \right).
    \label{eq:overall}
\end{multline}
The gating network and the context encoder together provide two complementary forms of environmental conditioning. The context encoder embeds scene information into every frame feature before temporal modeling, while the gating network translates the same environmental signals into expert routing weights that modulate the feed-forward transformation applied at each layer. This dual conditioning mechanism allows the model to adapt both its feature representations and its computational pathway to the propagation regime of each input sample.

\subsection{Spatial Feature Extraction}

The spatial encoder independently processes each pilot observation frame $\tilde{\mathbf{P}}_t \in \mathbb{R}^{2 \times K \times S_w}$. The primary objective of this module is to compress the two-dimensional spectral-spatial structure of the compressed pilot observations into a compact, fixed-length vector prior to the initiation of temporal modeling. The encoder comprises three convolutional layers, a squeeze-and-excitation recalibration block, and a global pooling stage, which are applied in sequence.

\subsubsection{Asymmetric Convolutional Layers}

The feature map $\tilde{\mathbf{P}}_t$ exhibits a well-defined physical structure. The height dimension ($K = 60$) represents the subcarrier positions and encodes the correlation within the frequency domain, whereas the width dimension ($S_w = 32$) indicates the wide-beam codewords and encodes the sparsity within the angular domain. The standard square convolution kernel treats the two axes symmetrically, which is inconsistent with their actual physical characteristics. Consequently, the encoder sequentially applies three convolutional layers, with each layer targeting a specific correlation structure.

The first layer utilizes a $1 \times 5$ kernel along the width axis:
\begin{equation}
    \mathbf{F}^{(1)} = \text{ReLU}\!\left(
    \text{Conv}_{1 \times 5}(\tilde{\mathbf{P}}_t)\right)
    \in \mathbb{R}^{32 \times K \times S_w},
    \label{eq:conv1}
\end{equation}
where a zero-padding configuration of $(0, 2)$ preserves the original spatial dimensions. This layer captures the correlation among adjacent beam codewords, a property that stems from the angular continuity of the mmWave channel. Beams directed toward spatially proximate angles produce similar pilot responses, and the $1 \times 5$ kernel exploits this local angular structure without incorporating frequency information at this stage.

The second layer applies a $5 \times 1$ kernel along the height axis:
\begin{equation}
    \mathbf{F}^{(2)} = \text{ReLU}\!\left(
    \text{Conv}_{5 \times 1}(\mathbf{F}^{(1)})\right)
    \in \mathbb{R}^{64 \times K \times S_w}.
    \label{eq:conv2}
\end{equation}
This layer extracts the correlation within the frequency domain across adjacent subcarriers. In the compressed pilot representation, the frequency selectivity of the channel manifests as smooth variations along the subcarrier axis. The span of the kernel is specifically selected to cover approximately one coherence bandwidth at a subcarrier spacing of 120~kHz.

The third layer utilizes a $3 \times 3$ kernel to model the joint spatial-frequency interactions:
\begin{equation}
    \mathbf{F}^{(3)} = \text{ReLU}\!\left(
    \text{Conv}_{3 \times 3}(\mathbf{F}^{(2)})\right)
    \in \mathbb{R}^{128 \times K \times S_w}.
    \label{eq:conv3}
\end{equation}
At this stage, the angular and frequency structures have been encoded separately by the preceding layers. Subsequently, the $3 \times 3$ convolution fuses them into joint representations to capture the variation trend of the channel frequency response across different beam directions.
This three-layer asymmetric design processes the angular and frequency information in dedicated stages prior to their combination. When compared with a stack of $3 \times 3$ kernels containing an equivalent parameter count, this architecture mitigates the risk of conflating physically distinct correlations in the early layers and yields intermediate representations that exhibit higher interpretability concerning the physics of the channel.

\subsubsection{Squeeze-and-Excitation Recalibration}

Following the three convolutional layers, the feature map $\mathbf{F}^{(3)} \in \mathbb{R}^{128 \times K \times S_w}$ comprises 128 channels with varying degrees of informativeness. Under NLOS conditions, or when the UE approaches the angular boundary of the coverage area, only a subset of channels carries discriminative beam information. The remaining channels respond primarily to noise or weak multipath components. The SE block performs a channel-wise recalibration to suppress these uninformative channels prior to the spatial compression of the feature map.
Initially, the SE block reduces the spatial dimensions through global average pooling:
\begin{equation}
    \mathbf{u} = \frac{1}{K \cdot S_w}
    \sum_{k,j} \mathbf{F}^{(3)}_{:,k,j} \in \mathbb{R}^{128},
    \label{eq:gap}
\end{equation}
Subsequently, the vector $\mathbf{u}$ is passed through a two-layer bottleneck network to derive a channel attention vector:
\begin{equation}
    \mathbf{g} = \sigma\!\left(\mathbf{W}_2\,
    \text{ReLU}(\mathbf{W}_1\,\mathbf{u})\right) \in \mathbb{R}^{128},
    \label{eq:se}
\end{equation}
where $\mathbf{W}_1 \in \mathbb{R}^{8 \times 128}$ and $\mathbf{W}_2 \in \mathbb{R}^{128 \times 8}$ constitute a bottleneck with a reduction ratio of $r = 16$, and $\sigma(\cdot)$ denotes the sigmoid function. The recalibrated feature map is formulated as:
\begin{equation}
    \mathbf{F}^{\text{se}} = \mathbf{F}^{(3)} \odot
    \mathbf{g}\!\left[\,:,\,\text{None},\,\text{None}\,\right],
    \label{eq:se_out}
\end{equation}
where $\odot$ denotes element-wise multiplication broadcast across the spatial axes. The inclusion of the SE block introduces merely $2 \times (128 \times 8) = 2048$ parameters, which represents a negligible overhead relative to the convolutional layers. Nevertheless, it provides a sample-adaptive mechanism that directs the attention of the network toward the channels that are most relevant to the current propagation conditions.

\subsubsection{Pooling and Projection}

The recalibrated feature map $\mathbf{F}^{\text{se}}$ is initially downsampled by a $2 \times 2$ max-pooling layer, which halves both spatial dimensions. Subsequently, an adaptive average pooling operation collapses the remaining spatial dimensions to $1 \times 1$ irrespective of the input resolution, yielding a 128-dimensional channel descriptor. A linear projection then maps this descriptor to the 256-dimensional frame embedding:
\begin{equation}
    \mathbf{z}_t^{\text{cnn}} = \mathbf{W}_{\text{fc}}\,
    \text{GAP}\!\left(\text{MaxPool}(\mathbf{F}^{\text{se}})\right)
    \in \mathbb{R}^{256}.
    \label{eq:cnn_out}
\end{equation}
The utilization of adaptive average pooling as the final spatial reduction step renders the encoder insensitive to the specific values of $K$ and $S_w$. Consequently, this architecture can accommodate variations in the size of the codebook or the subcarrier count without necessitating any structural modifications.

\subsection{Context Encoding and Scenario-Aware Gating}

The environmental context is incorporated into the model through two dedicated modules: a context encoder and a learnable gating network. Although both modules process identical inputs, specifically the scenario label $s$ and the normalized speed $\bar{v}$, they operate at different stages of the forward pass and fulfill distinct objectives. The context encoder embeds the environmental information into the representation of every frame prior to temporal modeling. Concurrently, the gating network translates identical signals into expert routing weights that govern the feed-forward computation within the GPT-2 backbone. This architectural separation ensures that the environmental conditioning acts on both the feature space and the computational pathway of the model.

\subsubsection{Context Encoder}

The context encoder maps the environmental pair $(s_t, \bar{v}_t)$ of each frame to a 32-dimensional embedding at every time step. The two scalars are initially concatenated into a vector and subsequently projected via a linear transformation:
\begin{equation}
    \mathbf{z}_t^{\text{ctx}} = \mathbf{W}_c
    \begin{bmatrix} s_t \\ \bar{v}_t \end{bmatrix}
    \in \mathbb{R}^{32}, \quad t = 1, \ldots, T,
    \label{eq:ctx}
\end{equation}
where $\mathbf{W}_c \in \mathbb{R}^{32 \times 2}$ represents a learned projection matrix. The output sequence $\mathbf{Z}^{\text{ctx}} \in \mathbb{R}^{B \times T \times 32}$ is concatenated with the CNN frame features $\mathbf{Z}^{\text{cnn}} \in \mathbb{R}^{B \times T \times 256}$ along the feature axis. Subsequently, the resulting 288-dimensional representation is projected to the hidden dimension of GPT-2 ($d = 768$) through a single linear layer, as formulated in~\eqref{eq:proj}.

The pilot tensor $\mathbf{X}$ conflates the propagation differences of LOS and NLOS environments, as well as speed-induced Doppler effects, into a single compressed observation. Under degraded channel conditions, such as severe shadowing or rapid movement of the UE, the signal-to-noise ratio of the pilot decreases, rendering these physical factors difficult to isolate solely from the observation. However, the context encoder provides a critical capability that the CNN encoder cannot achieve independently. By injecting $s_t$ and $\bar{v}_t$ as explicit features at the frame level, the context encoder supplies a reliable conditioning signal that maintains accuracy regardless of the quality of the pilot. Furthermore, this operation incurs a negligible computational cost.

\subsubsection{Scenario-Aware Gating Network}
\label{sec:gate}
The gating network determines the combination of the four expert networks within each MoE-FFN layer for a specific input sample. The input to this network consists of the environmental context of the final observed frame, $(s_T, \bar{v}_T)$. This design choice reflects the objective of the prediction task: the model needs to forecast the optimal beam at slot $T+1$, and the channel conditions at the final observed slot $T$ offer the closest available approximation to the conditions at the target slot.

The gating network comprises a two-layer MLP utilizing a ReLU activation function between the layers and a softmax output:
\begin{equation}
    \mathbf{w} = \text{Softmax}\!\left(
    \mathbf{W}_g^{(2)}\,\text{ReLU}\!\left(
    \mathbf{W}_g^{(1)}
    \begin{bmatrix} s_T \\ \bar{v}_T \end{bmatrix}
    \right)\right) \in \mathbb{R}^{4},
    \label{eq:gate}
\end{equation}
where $\mathbf{W}_g^{(1)} \in \mathbb{R}^{32 \times 2}$ and $\mathbf{W}_g^{(2)} \in \mathbb{R}^{4 \times 32}$ denote the weight matrices. The four output weights $\{w_i\}_{i=1}^{4}$ correspond to four propagation sub-conditions defined by the Cartesian product of the scenario type and the speed regime: LOS low-speed, LOS high-speed, NLOS low-speed, and NLOS high-speed. The threshold for the normalized speed is established at $\bar{v} = 0.5$, which corresponds to 60~km/h in physical units. The weight vector $\mathbf{w}$ is computed once per sample and shared across all $T$ time steps as well as all MoE layers. Consequently, the routing decision reflects a sample-level environmental condition rather than a token-level feature variation.

The architecture of the gating network diverges from the standard token-level MoE routing in two primary aspects. The routing signal is derived from explicit physical descriptors. This characteristic renders the assignment of experts directly interpretable. Furthermore, since the routing weights remain fixed for all frames within a sequence, the gating network avoids introducing load imbalance across experts at the token level.

\subsection{Temporal Modeling with MoE-Enhanced GPT-2}

\subsubsection{Transfer of Pretrained GPT-2}

The temporal backbone of the proposed model is constructed upon the pretrained GPT-2 transformer. The motivation for this architectural choice relies on a structural analogy between natural language modeling and beam sequence prediction. The beam prediction task exhibits exactly the same sequential structure as natural language modeling: given $T$ historical pilot observations, the model must predict the optimal beam index at slot $T+1$. 

Rather than training a transformer from a random initialization, the pretrained weights of GPT-2 provide a strong prior concerning sequential dependencies. This prior effectively transfers to the domain of beam prediction following fine-tuning. Only the initial four transformer blocks are retained, and the subsequent layers are discarded:
\begin{equation}
    \bigl\{\text{Block}_\ell\bigr\}_{\ell=0}^{3}
    \subset \text{GPT-2},
    \label{eq:gpt2_trunc}
\end{equation}
This truncation reduces the total parameter count while preserving sufficient modeling capacity for the ten-frame input sequences utilized in this work. The ablation study presented in Section~\ref{sec:ablation} confirms that four layers reach the performance saturation point for this task. 
The sequence input to the backbone is formulated by adding the pretrained positional embeddings $\mathbf{W}_{pe}$ to the projected frame embeddings:
\begin{equation}
    \mathbf{h}_t^{(0)} = \mathbf{e}_t +
    \mathbf{W}_{pe}[t] \in \mathbb{R}^{768},
    \quad t = 1, \ldots, T,
    \label{eq:pos_emb}
\end{equation}
where $\mathbf{e}_t$ is defined in~\eqref{eq:proj} and the positional index $t$ utilizes zero-based numbering. The positional embeddings encode the temporal order of the frames and are retained from the pretrained checkpoint. This retention provides the backbone with an awareness of the relative frame positions, an attribute that would otherwise necessitate additional training to acquire.

\subsubsection{MoE Feed-Forward Replacement}

Each transformer block applies a two-step update to the hidden sequence. The first step comprises a multi-head self-attention (MHSA) sublayer equipped with a residual connection and layer normalization:
\begin{equation}
    \mathbf{h}^{(\ell)}_{\text{attn}} =
    \mathbf{h}^{(\ell-1)} +
    \text{MHSA}\!\left(\text{LN}\!\left(\mathbf{h}^{(\ell-1)}\right)\right),
    \label{eq:attn}
\end{equation}
where LN denotes layer normalization applied prior to the sublayer, following the Pre-LN convention of GPT-2, which improves gradient stability during fine-tuning. The self-attention mechanism operates causally, meaning that the hidden state at position $t$ attends only to positions $1$ through $t$, consistent with the autoregressive structure of the prediction task.

The second step involves the feed-forward sublayer. In the initial block ($\ell = 0$), the original FFN of GPT-2 is retained:
\begin{equation}
    \mathbf{h}^{(0)} =
    \mathbf{h}^{(0)}_{\text{attn}} +
    \text{FFN}\!\left(\text{LN}\!\left(
    \mathbf{h}^{(0)}_{\text{attn}}\right)\right).
    \label{eq:ffn_std}
\end{equation} This layer performs low-level feature transformations that remain largely scene-independent, and a shared parameter set is therefore appropriate. In the subsequent blocks ($\ell = 1, 2, 3$), the FFN is replaced by a MoE-FFN that receives the gating weight vector $\mathbf{w}$ computed by the gating network in~\eqref{eq:gate}:
\begin{equation}
    \mathbf{h}^{(\ell)} =
    \mathbf{h}^{(\ell)}_{\text{attn}} +
    \text{MoE-FFN}\!\left(\text{LN}\!\left(
    \mathbf{h}^{(\ell)}_{\text{attn}}\right),\, \mathbf{w}\right),
    \quad \ell = 1, 2, 3.
    \label{eq:ffn_moe}
\end{equation} 
Each MoE-FFN comprises four expert networks $\{E_i\}_{i=1}^{4}$, where every expert adopts the two-layer structure of the original GPT-2 FFN:
\begin{equation}
    E_i(\mathbf{h}) =
    \text{Dropout}\!\left(
    \mathbf{W}_i^{(2)}\,\text{GELU}\!\left(
    \mathbf{W}_i^{(1)}\mathbf{h}\right)\right),
    \label{eq:expert}
\end{equation}
with $\mathbf{W}_i^{(1)} \in \mathbb{R}^{3072 \times 768}$ and $\mathbf{W}_i^{(2)} \in \mathbb{R}^{768 \times 3072}$, matching the expansion ratio of four used in the original GPT-2 FFN. The routing mode applied to the MoE-FFN differs across the three training stages and at inference, as detailed in Section~\ref{sec:training}. During Stage~1, a hard assignment mask activates exactly one expert per sample based on the ground-truth scene label~\eqref{eq:hard_route}. During Stage~2, all four experts are activated under full soft routing to train the gating network:
\begin{equation}
    \text{MoE-FFN}(\mathbf{h}_t, \mathbf{w}) =
    \sum_{i=1}^{4} w_i \cdot E_i(\mathbf{h}_t).
    \label{eq:moe_soft}
\end{equation}
During Stage~3 and at inference, top-1 hard routing is applied: only the expert with the highest gating weight is activated:
\begin{equation}
    \text{MoE-FFN}(\mathbf{h}_t, \mathbf{w}) =
    E_{i^*}(\mathbf{h}_t), \quad
    i^* = \arg\max_{i \in \{1,2,3,4\}} w_i.
    \label{eq:moe_top1}
\end{equation}
Because $\mathbf{w}$ is fixed for all $T$ time steps within a sample, the routing decision is applied uniformly across the entire frame sequence. Top-1 hard routing at inference eliminates inter-expert interference in the output, reduces the feed-forward computation per layer to that of a single expert, and ensures that the model applies the most suitable scene-specific transformation without blending outputs from less relevant experts.

\subsubsection{Output Extraction}

Following the four transformer blocks, the hidden state at the final time step is extracted:
\begin{equation}
    \mathbf{h}_T =
    \mathbf{h}^{(3)}[:, -1, :] \in \mathbb{R}^{768}.
    \label{eq:last_token}
\end{equation}
The causal structure of the self-attention mechanism guarantees that $\mathbf{h}_T$ aggregates information from all preceding frames ($\mathbf{h}_1, \ldots, \mathbf{h}_{T-1}$). Therefore, no additional sequence pooling is required. A dropout layer with a rate of $p = 0.2$ is applied to $\mathbf{h}_T$ as a regularization measure prior to the computation of the final $C$-dimensional logits by the linear classifier:
\begin{equation}
    \hat{\mathbf{y}} = \mathbf{W}_{\text{cls}}\,
    \text{Dropout}(\mathbf{h}_T) \in \mathbb{R}^{C},
    \label{eq:logits}
\end{equation}
Subsequently, the predicted beam index is determined as $\hat{b}^* = \arg\max_c\, \hat{y}_c$.

\subsection{Three-Stage Training Strategy}
\label{sec:training}
Training the proposed architecture in an end-to-end manner from a random initialization proves ineffective for two primary reasons. First, the pretrained weights of GPT-2 encode useful sequential priors; optimizing all parameters simultaneously subjects these priors to destruction by large gradient updates during early training. More importantly, the specialization of the MoE experts remains fragile. Without a structured initialization phase, all experts receive similar gradient signals and converge to nearly identical functions, expert collapse occurs. Consequently, the MoE structure degenerates into a standard FFN with redundant parameters. To mitigate both issues, the training process is organized into three sequential stages, each characterized by a distinct set of trainable parameters and specific optimization objectives.

\subsubsection{Stage 1: Expert Pre-Training with Hard Routing}

In the first stage, all model parameters are jointly optimized from the pretrained GPT-2 initialization. A hard routing mechanism is used: before each forward pass, an expert assignment mask $\mathbf{m} \in \{0,1,2,3\}^{B}$ is constructed from the ground-truth scene labels and speed regimes of the current mini-batch, where the assignment rule follows the four-way partition defined in Section~\ref{sec:gate}. The mask is broadcast to all $T$ time steps and passed to each MoE-FFN layer, which activates only the assigned expert for each sample:
\begin{equation}
    \text{MoE-FFN}(\mathbf{h}_t,\, m_b) = E_{m_b}(\mathbf{h}_t),
    \quad b = 1, \ldots, B.
    \label{eq:hard_route}
\end{equation}
Under this scheme, the gradient of the classification loss with respect to Expert~$i$ flows only through samples assigned to that expert, so each expert receives a consistent scene-specific supervision signal throughout training. By the end of Stage~1, the four experts have developed distinct parameter spaces that reflect the different temporal dynamics of their respective propagation sub-conditions.

Note that the gating network is present in the model during Stage~1 but plays no role in routing: the expert mask takes priority in the MoE-FFN forward pass and the gate output is disregarded. This preserves the gating network as an uninitialized module that will be trained independently in Stage~2.

\subsubsection{Stage 2: Gating Network Alignment with Soft Routing}

The second stage loads the Stage~1 checkpoint and freezes all model parameters except the gating network. All expert weights, the CNN encoder, the context encoder, the feature projection layer, and the classifier are held fixed. The expert mask is removed, and the MoE-FFN reverts to dense soft routing, where all experts are activated:
\begin{equation}
    \text{MoE-FFN}(\mathbf{h}_t, \mathbf{w}) =
    \sum_{i=1}^{4} w_i \cdot E_i(\mathbf{h}_t).
\end{equation}
with routing weights supplied by the learnable gate:
\begin{equation}
    \mathbf{w} = \text{Gate}(s_T, \bar{v}_T) \in \mathbb{R}^4.
    \label{eq:gate_stage2}
\end{equation} 
The sole trainable component is the gating network, which has a parameter count of only $(2 \times 32) + (32 \times 4) = 192$.
The objective of this stage is to train the gate to produce routing weights that approximate the hard assignment of Stage~1 in a differentiable, continuous form. Because the expert parameters are frozen, the gate must learn to assign high weight to the expert that was pre-trained on the scene matching the current input. The cross-entropy loss on the beam classification task provides indirect supervision for this alignment: a correct gate assignment activates the most capable expert for the current scene, which in turn minimizes the prediction error. By the end of Stage~2, the gate weight distribution exhibits clear diagonal dominance, with each scene sub-condition receiving its highest weight on the corresponding pre-trained expert.

The higher learning rate compared to Stage~1 is appropriate, because only the small gating network is being updated and fast convergence is desirable.

\subsubsection{Stage~3: Top-1 Hard Fine-Tuning}

The third stage loads the Stage~2 checkpoint and performs targeted fine-tuning under top-1 hard routing and a hierarchical parameter freezing scheme. The routing mode switches from the full soft routing of Stage~2 to deterministic single-expert activation: for each forward pass, the expert index $i^* = \arg\max_i w_i$ is selected by the gating network, and only $E_{i^*}$ is evaluated. The gradient of the classification loss therefore flows exclusively through the selected expert, reinforcing scene-specific specialization rather than averaging gradients across all experts as in soft routing.

The frozen and trainable components are as follows. The entire GPT-2 self-attention mechanism and the CNN spatial encoder are frozen. Within each expert network, the first linear layer $\mathbf{W}_i^{(1)} \in \mathbb{R}^{3072 \times 768}$ is frozen and only the second linear layer $\mathbf{W}_i^{(2)} \in \mathbb{R}^{768 \times 3072}$ remains trainable. The gating network and the linear classifier are fully trainable.

This asymmetric freezing is motivated by the functional role of each layer. The first linear layer expands the hidden representation into the intermediate dimension and captures low-level scene-independent transformations that were established during Stage~1. The second linear layer projects back to the hidden dimension and performs the scene-specific output transformation, which benefits from further refinement under the hard-routing objective. Because top-1 routing directs the full gradient signal to a single expert per sample, the second layers of the selected experts receive stronger and more consistent scene-specific updates than they would under soft or sparse routing, where the gradient is divided among multiple experts.

The routing mode at inference is identical to that used in Stage~3 training, ensuring that the parameter distribution optimized during fine-tuning matches the activation pattern seen at deployment. This training-inference consistency is a key advantage of hard routing over soft routing, where training uses weighted blending but inference typically applies a different selection criterion.

The optimizer is AdamW with parameter-group-specific learning rates and a shared weight decay of $\lambda = 10^{-4}$:
\begin{equation}
    \eta_k =
    \begin{cases}
        2 \times 10^{-4} & \text{classifier and gating network,} \\
        5 \times 10^{-5} & \text{expert second layers.}
    \end{cases}
    \label{eq:lr_stage3}
\end{equation}
Training runs for up to 200 epochs with an early stopping patience of 10 epochs.

\section{Simulation Results}
\label{sec:simulation}
\subsection{Simulation Configuration}

\subsubsection{Dataset Generation}
The simulation data is generated by Quadriga, in compliance with the Urban Macro (UMa) scenario specifications defined in 3GPP TR 38.901. The operating frequency is 28 GHz (FR2) with a subcarrier spacing of 120 kHz. The base station (BS) is equipped with a 64-element uniform linear array (ULA), and each user equipment (UE) is equipped with a single omnidirectional antenna. A total of 64,000 UE samples (sequences) are randomly generated within a distance of 30 to 100 m and an angle range of $\pm60^{\circ}$ from the base station. The user mobility speed is uniformly distributed between 0 and 120 km/h to simulate realistic mobile communication scenarios. The channel sampling interval is set to 10 ms. Each sample sequence contains 11 consecutive slots, where the first 10 slots serve as the observation input and the 11th slot acts as the prediction target. Zadoff-Chu (ZC) sequences are adopted as uplink pilot signals, and a DFT-based wide beam codebook is utilized for beam training. The dataset is randomly partitioned into training, validation, and test sets with a ratio of 70:15:15. Detailed model parameters are summarized in Table~\ref{tab:params}.

\begin{table}[htbp]
\centering
\caption{Key Simulation and Dataset Parameters}
\label{tab:params}
\small
\begin{tabular}{lc}
\toprule
Parameter & Value \\
\midrule
Carrier frequency           & 28~GHz \\
Subcarrier spacing          & 120~kHz \\
Active subcarriers ($K$)    & 60 \\
Antenna configuration (BS)  & 64-element ULA \\
Antenna configuration (UE)  & Single omnidirectional \\
RF chains ($N_{rf}$)        & 8 \\
Wide-beam codebook size ($S_w$) & 32 \\
UE transmit power ($p_u$)   & 200~mW \\
Noise power ($\sigma_n^2$)  & $-123$~dBm \\
Distance range              & 30--100~m \\
Angle range                 & $[-60^{\circ},\,60^{\circ}]$ \\
UE speed range              & 0--120~km/h \\
Sampling interval           & 10~ms \\
Sequence length (Input + Target) & $10 + 1$ \\
Dataset scenario            & 3GPP TR~38.901 UMa \\
Total UE samples            & 64{,}000 \\
Dataset split (Train:Val:Test) & 70\%:15\%:15\% \\
\bottomrule
\end{tabular}
\end{table}

\subsubsection{Hyperparameters and Training Settings}
The model architecture comprises a 3-layer CNN encoder integrated with a squeeze-and-excitation (SE) module, a context encoder, and a 4-layer lightweight GPT-2 backbone. The FFN in the last three layers of the GPT-2 backbone is replaced by a 4-expert mixture-of-experts (MoE) module. The training process adopts the proposed three-stage strategy. During the final fine-tuning phase, the CNN encoder and self-attention layers are hierarchically frozen to focus on optimizing the top-layer MoE experts and the gating network. The model is trained and fine-tuned using the AdamW optimizer paired with a cosine annealing warm restart scheduler. An early stopping mechanism with a patience of 10 epochs is introduced to prevent model overfitting. The detailed model parameters are summarized in Table~\ref{tab:model_architecture}.
\begin{table}[h]
\centering
\caption{Model Architecture and Parameters}
\label{tab:model_architecture}
\setlength{\tabcolsep}{1pt} 
\begin{tabular}{ccc}
\toprule
Module & Layer & Parameter \\
\midrule
\multirow{9}{*}{\shortstack{Beam CNN\\Encoder}} 
 & Convolution & $f_i = 2, f_o = 32, (1, 5)$ \\
 & ReLU activation & $f_i = 32, f_o = 32$ \\
 & Convolution & $f_i = 32, f_o = 64, (5, 1)$ \\
 & ReLU activation & $f_i = 64, f_o = 64$ \\
 & Convolution & $f_i = 64, f_o = 128, (3, 3)$ \\
 & ReLU activation & $f_i = 128, f_o = 128$ \\
 & SE Block & $f_i = 128, f_o = 128, \text{reduction} = 16$ \\
 & Pooling & max-pooling $\to$ adaptive-avg-pooling \\
 & FC & $f_i = 128, f_o = 256$ \\
\midrule
\multirow{2}{*}{\shortstack{Context\\Encoder}} 
 & FC & $f_i = 2, f_o = 32$ \\
 & Feature Projection & $f_i = 288, f_o = 768$ \\
\midrule
\multirow{2}{*}{\shortstack{Learnable\\Gate}} 
 & FC + ReLU & $f_i = 2, f_o = 32$ \\
 & FC + Softmax & $f_i = 32, f_o = 4$ \\
\midrule
\multirow{4}{*}{\shortstack{GPT-2 \\ \& MoE Blocks\\(4 layers)}} 
 & Transformer Self-Attn & $f_i = 768, f_o = 768$ \\
 & \multirow{2}{*}{Expert FFN ($\times 4$)} & $f_i = 768, f_o = 3072, \text{GELU}$ \\
 & & $f_i = 3072, f_o = 768, \text{dropout} = 0.1$ \\
\midrule
\multirow{2}{*}{\shortstack{Output\\Module}} 
 & Dropout & $f_i = 768, f_o = 768, \text{dropout} = 0.2$ \\
 & FC (Classifier) & $f_i = 768, f_o = 32$ \\
\bottomrule
\end{tabular}
\end{table}

\subsubsection{Metrics}

Several core evaluation metrics are employed to comprehensively assess the performance of the proposed beam prediction model. These metrics include the Top-K accuracy, the conditional accuracy under dynamic beam switching, and the real-time inference latency. 
\paragraph{Top-K Accuracy}
The core metric for beam prediction is the Top-K accuracy, which measures the proportion of test samples where the true beam index falls into the top K predicted candidate set. It is formally defined as:
\begin{equation}
    \text{Acc}_K = \frac{1}{N} \sum_{i=1}^{N} \mathbb{I}\left(y_i \in \hat{\mathcal{Y}}_{i,K}\right)
\end{equation}
where $N$ is the total number of test samples, $y_i$ represents the true beam index of the $i$-th sample, and $\hat{\mathcal{Y}}_{i,K}$ denotes the set of the top K predicted beam indices sorted by their prediction probabilities. The indicator function $\mathbb{I}(\cdot)$ returns 1 if the condition is met and 0 otherwise. This experiment reports both Top-1 and Top-3 accuracies to evaluate the absolute precision of the model and the reliability of providing alternative robust beam candidates, respectively.
\paragraph{Accuracy in Highly Dynamic Scenarios}
In practical millimeter-wave communication systems, sudden channel variations severely challenge link reliability and model robustness. Leveraging the inherent temporal continuity of optimal beam sequences, we introduce transition accuracy as a critical supplementary evaluation metric beyond the conventional overall accuracy. The transition accuracy is calculated exclusively on a subset of samples where the optimal beam index changes between adjacent time steps, defined as $b_T^* \neq b_{T+1}^* $. This metric rigorously evaluates the ability of the model to capture rapid spatial dynamics and predict sudden beam switches rather than passively repeating historical predictions, thereby reflecting model robustness. 

\paragraph{Real-Time Inference Latency}
Comprehensive inference latency tests are conducted to verify the feasibility of deploying the proposed MoE-enhanced GPT-2 model in real-time online systems. In addition to the average processing time per sample, the 99th percentile latency is adopted. The purpose of this evaluation is not to compare inference speed across architectures, but to confirm that the proposed model meets the latency requirements of slot-level beam management in 3GPP NR systems, where the coherence interval is 10~ms.

All experiments are conducted on a server equipped with an NVIDIA H100 PCIe 80GB GPU. Inference latency is evaluated in single-sample mode with a batch size of 1, where the input tensor has shape (1, 10, 2, 60, 32). The model operates in FP16 half-precision. Prior to formal measurement, each model undergoes 20 warm-up forward passes to eliminate cold-start overhead caused by GPU kernel initialization. Latency statistics are then collected over 1000 consecutive inferences.

\subsection{Baseline Comparison}

\begin{table*}[htbp]
\centering
\caption{Comparison Among Different Schemes}
\label{tab:baseline}
\renewcommand{\arraystretch}{1.2}
\setlength{\tabcolsep}{3pt}
\begin{tabular}{lccccccccc}
\toprule
\multirow{2}{*}{Model Variant} & \multirow{2}{*}{Top-1 (\%)} & \multirow{2}{*}{Top-3 (\%)} & \multirow{2}{*}{Transition Acc (\%)} & \multirow{2}{*}{Avg Latency (ms)} & \multirow{2}{*}{P99 Latency (ms)} & \multicolumn{4}{c}{Scene-wise Top-1 Accuracy (\%)} \\
\cmidrule(lr){7-10}
 & & & & & & LOS-L & LOS-H & NLOS-L & NLOS-H \\
\midrule
Full Model (Ours) & \textbf{94.88} & \textbf{99.83} & \textbf{80.62} & 0.52 & 0.52 & \textbf{93.30} & \textbf{95.21} & \textbf{94.83} & \textbf{96.24} \\
CNN           & 81.15 & 99.68 & 29.22 & \textbf{0.13} & \textbf{0.14} & 82.07 & 78.03 & 84.48 & 79.94 \\
CNN+LSTM      & 92.66 & 99.81 & 67.59 & 0.32 & 0.33 & 91.59 & 92.01 & 92.94 & 92.31 \\
CNN+GPT2      & 92.55 & 99.80 & 71.07 & 0.32 & 0.32 & 91.14 & 92.79 & 92.56 & 93.75 \\
\bottomrule
\end{tabular}
\end{table*}

The table \ref{tab:baseline} presents a performance comparison between the proposed model and three baseline methods, highlighting its superior performance. The baseline methods comprise deep learning approaches with different architectures, specifically a CNN, a CNN+LSTM model, and a CNN+GPT-2 model. Table~\ref{tab:baseline} evaluates the overall system-level gain of the proposed framework relative to architectures of varying complexity. The individual contributions of the MoE mechanism, context-aware gating, and three-stage training are further decomposed in the ablation study of Section~\ref{sec:ablation}. All methods adopt identical relevant settings without affecting model training to ensure a fair comparison.
\subsubsection{CNN}
The traditional Convolutional Neural Network (CNN) captures millimeter-wave channel information. The integration of the Squeeze-and-Excitation (SE) module further assists the model in understanding the spatial-frequency correlations of the channel to extract relevant features. The CNN baseline extracts spatial features from each pilot observation frame independently. With the SE module, the encoder learns to weight informative frequency-beam correlations and suppress redundant channels. Its average latency of $0.13$ ms reflects the simplicity of the architecture. However, the model has no mechanism to relate observations across time. Each frame is processed in isolation, and the final prediction is made from a single-frame representation. At beam switching instants, the optimal beam changes precisely because of temporal channel dynamics, which a frame-independent model cannot observe. The transition accuracy of $29.22\%$ is therefore close to what a random or persistence predictor would achieve, confirming that spatial feature quality alone is insufficient for this task.
\subsubsection{CNN+LSTM}
The addition of a Long Short-Term Memory (LSTM) network introduces temporal prediction capabilities, enabling the model to utilize historical information. This increases the Top-1 accuracy to $92.66\%$ and the transition moment accuracy to $67.59\%$. Its lightweight architecture also maintains a low latency of approximately $0.32$ ms. Nevertheless, the LSTM has a weak capacity for long-sequence dependencies, and its sequential processing nature causes inevitable memory degradation, which limits the prediction accuracy at transition moments. The remaining gap relative to the proposed model is attributable to the structural limitations of the LSTM. The hidden state of LSTM is updated sequentially, so information from early frames must pass through multiple gating operations before influencing the final prediction. Under high mobility, where channel conditions can shift substantially within ten frames, this propagation path introduces information loss that self-attention could avoid. Moreover, the LSTM has no mechanism to modulate its behavior based on the propagation environment. It applies the same recurrent weights regardless of the channel state, which limits its ability to adapt to the distinct temporal dynamics of each scene.
\subsubsection{CNN+GPT2}
The CNN+GPT-2 model retains the first four layers of the GPT-2 architecture, matching the backbone depth of the proposed model while omitting the MoE feed-forward replacement and the context-aware gating mechanism. This variant is trained end-to-end without the three-stage strategy. Its Top-1 accuracy reaches $92.55\%$ and its transition accuracy improves to $71.07\%$ over CNN+LSTM, at an average inference latency of $0.32$~ms. The gain over CNN+LSTM at beam switching instants is attributable to the self-attention mechanism. By attending to all ten frames simultaneously, the transformer backbone retains information from early slots without the information loss that accumulates through sequential hidden-state propagation in an LSTM. The two models share the same CNN spatial encoder, which isolates the temporal backbone as the source of this improvement.
The accuracy gap relative to the proposed model, however, reveals the limitation of a scene-agnostic transformer. Without expert routing conditioned on propagation descriptors, the same set of feed-forward parameters must serve all four propagation sub-conditions simultaneously. The parameter capacity that could otherwise be allocated to scene-specific temporal patterns is diluted across heterogeneous channel regimes, constraining the representational specialization achievable within each individual condition. This comparison directly motivates the scene-conditioned MoE-FFN design and, separately, the three-stage training strategy: both the architectural conditioning and the structured initialization are necessary to realize the full accuracy gain over a plain GPT-2 backbone.
\subsubsection{Full Model}
The proposed architecture integrates scene-conditioned MoE feed-forward layers into a four-layer GPT-2 backbone and conditions the expert routing on explicit environmental context. This combination addresses the two limitations identified in the CNN+GPT-2 baseline: the absence of scene-specific parameter specialization and the uniform treatment of all propagation regimes within a single FFN. The transition accuracy of $80.62\%$ represents a gain of $9.55$ percentage points over CNN+GPT-2, confirming that physically grounded expert routing substantially improves the prediction of abrupt spatial transitions. The decomposition of this gain into contributions from the MoE mechanism, context encoder, and training strategy is presented in Table~\ref{tab:ablation}.
In terms of inference latency, the adoption of top-1 hard routing in Stage~3 means that only one expert is activated per forward pass, so the effective feed-forward computation per MoE-FFN layer is equivalent to that of a single standard FFN. The average latency of $0.52$~ms is nonetheless higher than that of CNN+GPT-2 ($0.32$~ms), because all four expert parameter matrices must reside in GPU memory and are subject to memory bandwidth overhead during weight loading, irrespective of how many experts are arithmetically activated. This overhead is inherent to the MoE parameter structure and is independent of the routing strategy. Critically, $0.52$~ms remains well within the $10$~ms channel coherence interval of the 3GPP NR frame structure, confirming that the model satisfies the latency requirements of slot-level beam management. The primary purpose of reporting latency is to verify deployment feasibility, not to claim speed superiority over architecturally simpler baselines.
Across the four scene subsets, the proposed model achieves the highest accuracy in all categories, with particularly strong gains in high-speed scenarios where channel dynamics are most severe. This pattern is consistent with the design intent: the gating network assigns the highest weight to the speed-matched expert under high-mobility conditions, and top-1 hard routing ensures that this expert alone handles the prediction, concentrating model capacity precisely where it is most needed.

\subsection{Ablation Studies}
\label{sec:ablation}

\begin{table*}[htbp]
\centering
\caption{Ablation Studies}
\label{tab:ablation}
\renewcommand{\arraystretch}{1.2}
\setlength{\tabcolsep}{3pt}
\begin{tabular}{lccccccccc}
\toprule
\multirow{2}{*}{Model Variant} & \multirow{2}{*}{Top-1 (\%)} & \multirow{2}{*}{Top-3 (\%)} & \multirow{2}{*}{Transition Acc (\%)} & \multirow{2}{*}{Avg Latency (ms)} & \multirow{2}{*}{P99 Latency (ms)} & \multicolumn{4}{c}{Scene-wise Top-1 Accuracy (\%)} \\
\cmidrule(lr){7-10}
 & & & & & & LOS-L & LOS-H & NLOS-L & NLOS-H \\
\midrule
Full Model (Ours) & \textbf{94.88} & \textbf{99.83} & \textbf{80.62} & 0.52 & 0.52 & \textbf{93.30} & \textbf{95.21} & \textbf{94.83} & \textbf{96.24} \\
w/o MoE           & 91.17 & 99.74 & 67.00 & \textbf{0.33} & \textbf{0.33} & 88.73 & 91.56 & 91.42 & 93.03 \\
w/o Context       & 90.21 & 99.77 & 67.99 & 0.55 & 0.55 & 87.51 & 91.60 & 90.08 & 91.77 \\
w/o SE            & 79.77 & 99.59 & 42.45 & 0.54 & 0.54 & 77.58 & 80.70 & 79.35 & 81.50 \\
End-to-End        & 93.04 & 99.46 & 72.96 & 0.55 & 0.55 & 91.71 & 93.00 & 93.40 & 94.09 \\
\bottomrule
\end{tabular}
\end{table*}

\begin{figure*}[t]
    \centering
    \begin{minipage}{0.48\linewidth}
        \centering
        \includegraphics[width=\linewidth]{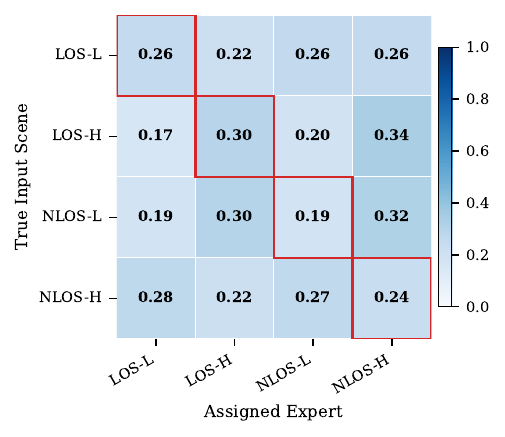}
        \caption{Gate weight heatmap of the end-to-end trained model, 
where uniform expert assignment indicates expert collapse.}
        \label{fig:e2e}
    \end{minipage}
    \hfill
    \begin{minipage}{0.48\linewidth}
        \centering
        \includegraphics[width=\linewidth]{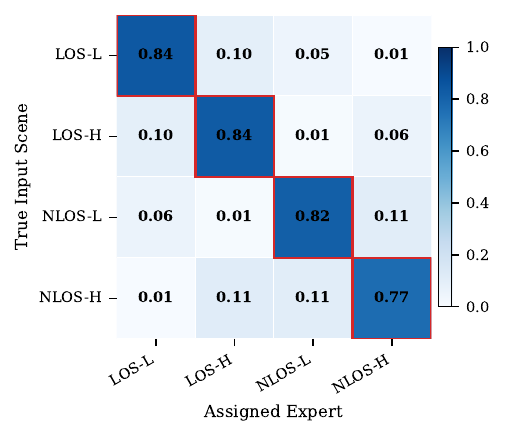}
        \caption{Gate weight heatmap of the proposed three-stage trained 
model, where diagonal dominance confirms scene-aligned expert routing.}
        \label{fig:our gate}
    \end{minipage}
\end{figure*}

\subsubsection{Component Ablation}

To rigorously assess the individual contributions of key components within the proposed architecture, a comprehensive suite of ablation studies was conducted. By benchmarking the performance of the full model against the following three distinct variants, we demonstrate the indispensable roles of the Mixture-of-Experts (MoE) mechanism, environmental context-awareness, the selection of the temporal backbone, and our optimized training strategies.

\paragraph{Ablation 1 (w/o MoE)}Replace the MoE layer in the GPT-2 backbone with a standard FFN to assess the impact of dynamic expert routing on performance.

\paragraph{Ablation 2 (w/o Context)} Remove the learnable context gating and encoder, assigning expert weights uniformly to evaluate the importance of scenario-aware routing.

\paragraph{Ablation 3 (w/o SE)} The Squeeze-and-Excitation (SE) block is removed from the Beam CNN Encoder to verify the effectiveness of channel-wise attention for spatial feature extraction.

Table \ref{tab:ablation} presents a performance comparison across various components of the proposed model. Compared to three distinct variants, the proposed architecture achieves the superior performance across all evaluation metrics. While maintaining a Top-1 prediction accuracy of $94.88\%$ and an average inference latency of $0.52$ms, the model reaches an accuracy of $80.62\%$ for beam transition moments, which exceeds that of other variants by more than $7\%$.

Reverting the replacement of standard FFN layers in the GPT-2 backbone causes the prediction accuracy for transition moments to decrease by over $13$ percentage points. This indicates that dynamic expert routing better captures abrupt changes in channel scenario information. The integration of environmental context information provides external guidance for the learning of expert routing, which assists in stabilizing model performance. Consequently, removing this component leads to a general decline in prediction accuracy. The most severe performance degradation occurs when the Squeeze-and-Excitation (SE) module is removed, as the Top-1 prediction accuracy falls to $79.77\%$ and the transition moment accuracy drops significantly to $42.45\%$. This validates the necessity of spatial feature extraction for temporal prediction. 

The ablation results also reveal an important interaction among the three core components. The SE block, the context-aware gating, and the MoE layers are not independently additive; they form a mutually reinforcing pipeline. The SE block provides channel-wise attention that sharpens the spatial representation fed into the temporal backbone, giving the gating network a more discriminative feature basis for routing decisions. The context encoder then supplies explicit environmental signals that stabilize the routing independently of the pilot observation quality, which can degrade under high mobility or heavy NLOS scattering. Without both of these upstream components in place, the MoE layers receive either poorly discriminated spatial features or ambiguous routing signals, and the expert specialization collapses or overfits to dominant scene patterns in the training set. In other words, using MoE alone without the SE block and context encoder does not improve over a standard model and can in fact hurt generalization, as the additional expert capacity amplifies rather than compensates for upstream representation deficiencies.

\subsubsection{Training Strategy Analysis}
\begin{figure*}[t]
    \centering
    \begin{minipage}{0.48\linewidth}
        \centering
        \includegraphics[width=1\linewidth]{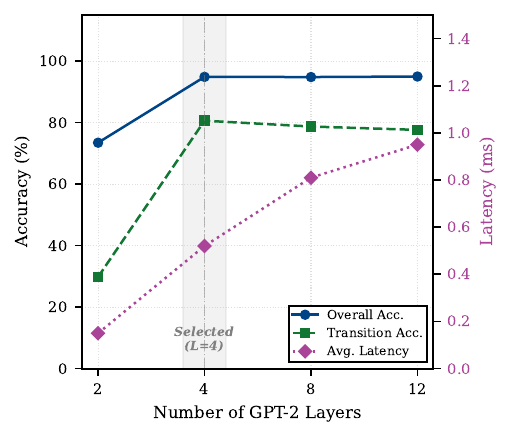}
        \caption{Impact of GPT-2 Depth}
        \label{fig:gpt2}
    \end{minipage}
    \hfill
    \begin{minipage}{0.48\linewidth}
        \centering
        \includegraphics[width=1\linewidth]{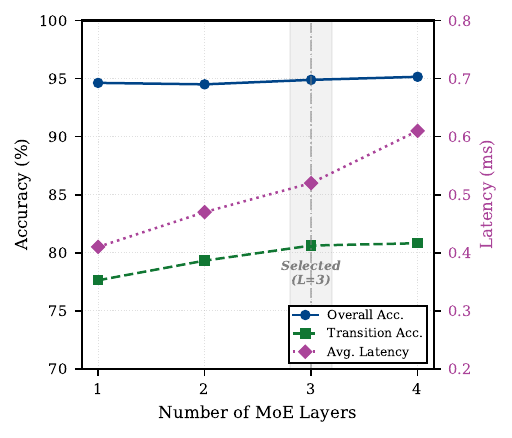}
        \caption{Impact of MoE Replacement Layers}
        \label{fig:moelayer}
    \end{minipage}
\end{figure*}
For the proposed three-stage training strategy, we introduce a novel variant. This model discards the proposed three-stage hierarchical training strategy and trains the model end-to-end from the pretrained GPT-2 initialization without the structured three-stage curriculum, which verifies the necessity of this curriculum learning strategy in preventing expert collapse.

As shown in table \ref{tab:ablation}, adopting end-to-end training instead of the three-stage strategy maintains relatively stable overall accuracy and latency but severely compromises the transition moment prediction accuracy by more than $10\%$. Fig \ref{fig:e2e} and Fig \ref{fig:our gate} demonstrate that end-to-end training induces expert collapse, the gating weights become nearly uniform across all four experts regardless of the input scene. Specifically, the variation in expert weight distribution for each scenario is constrained within $10\%$, which means the MoE structure degenerates into a standard FFN with redundant parameters. Therefore, the multi-stage training strategy is indispensable for preventing expert collapse and ensuring that specific network parameters effectively handle complex temporal transitions.
The ablation results in Table~\ref{tab:ablation} further decompose the sources of performance gain. Compared to the w/o~MoE variant, the MoE-FFN mechanism alone improves transition accuracy by $13.62$ percentage points, confirming the central role of scene-conditioned expert routing. Compared to the w/o~Context variant, the addition of the context encoder yields a gain of $12.63$ percentage points, reflecting the importance of explicit environmental conditioning. The three-stage training strategy contributes approximately $7.66$ percentage points over the end-to-end variant. These three comparisons share the same full model as the upper bound but differ in their respective lower bounds, and therefore cannot be summed to reconstruct the $9.55$ percentage point gain over CNN+GPT-2, which reflects the combined benefit of all three components relative to a single-component baseline trained end-to-end.

\subsection{Architecture Design Analysis}
\subsubsection{Impact of GPT-2 Depth}

To evaluate the impact of network depth, four models are compared retaining the first two, four, eight, and twelve layers of the GPT-2 architecture. For the 8-layer and 12-layer variants, we freeze the first 4 and 8 layers respectively during training to mitigate overfitting. Except for the 2-layer variant which replaces both FFN layers with Mixture-of-Experts layers, other models substitute the feed-forward layers in the final four layers. All other component configurations remain identical.

Results in Fig \ref{fig:gpt2} show that transition moment accuracy exhibits non-linear growth as GPT-2 layers increase. The 2-layer model exhibits extremely poor performance in terms of transition accuracy. When four layers are retained, the transition accuracy increases, indicating that greater GPT-2 depth strengthens temporal prediction capabilities, the 4-layer Transformer structure effectively captures the temporal evolution patterns of beams. Adding extra layers further degrades the accuracy of the model beyond the four-layer configuration, proving the model reaches diminishing returns and performance saturation. Beyond four layers, gains are marginal as overfitting occurs, while inference latency increases by approximately $0.2$ to $0.4$ ms for every four additional layers. This latency makes completing beam handovers difficult within channel coherence times in high-mobility scenarios. Compared to the 2-layer structure, the 4-layer GPT-2 model improves performance while maintaining a low average latency, achieving an optimal balance between communication performance and computational resources.

\subsubsection{Impact of MoE Replacement Layers}

Building upon the four-layer GPT-2 backbone, we evaluate Mixture-of-Experts substitution layers. The lower layers are responsible for extracting generic low-level features, while the higher layers are tasked with modeling task-specific high-level semantic features, which are suited for processing by expert modules. Similar to GPT-2 layer retention, increasing substitution layers enhances expert capacity to model temporal beam variations but concurrently increases inference latency. As shown in Fig \ref{fig:moelayer}, each additional MoE layer substitution increases average inference latency by $0.05$ to $0.10$~ms, while the marginal improvement in transition accuracy diminishes progressively, indicating that the model approaches its fitting capacity saturation.

We ultimately replace FFN layers in the final three GPT-2 layers with Mixture-of-Experts layers. Compared to replacing only the last two layers, the proposed model improves transition accuracy to over $80\%$ while maintaining an average inference latency of $0.52$ ms. Compared to replacing FFNs in all layers, the proposed model sacrifices only about $0.2\%$ in performance but reduces latency by nearly $0.1$ ms, providing a computational margin and safety redundancy.

\section{CONCLUSION}
\label{sec:conclusion}
This paper presented CAT-MoEformer, a proactive beam prediction framework integrating a pretrained GPT-2 temporal backbone with scene-conditioned MoE feed-forward layers for mmWave beam management. Three design elements jointly address limitations of existing methods. The asymmetric convolutional encoder with squeeze-and-excitation extracts spatially and spectrally discriminative features from compressed pilot observations, providing a reliable representation for both the temporal backbone and gating network. The scene-conditioned MoE routes each sample to the most suitable expert using explicit propagation descriptors, allowing specialization according to LOS/NLOS regimes and UE mobility. The three-stage training strategy avoids expert collapse under end-to-end soft routing by first establishing scene-specific expert specialization via hard assignment using ground-truth scene labels, then aligning the gating network under full soft routing, and finally refining expert output layers and the classifier under top-1 hard routing. Transitioning from soft routing in Stage~2 to hard routing in Stage~3 ensures identical training and inference modes, eliminating the mismatch that continuous soft routing would cause.

Simulation results on 3GPP TR~38.901 Urban Macro channel simulations showed consistent accuracy gains over CNN, CNN+LSTM, and CNN+GPT-2 baselines across all metrics, with the largest improvement at beam switching instants where transition accuracy of $80.62\%$ exceeds the CNN+GPT-2 baseline by $9.55$ points, indicating scene-conditioned expert routing enhances abrupt spatial transition prediction. Gate weight heatmaps confirmed that the three-stage strategy produces well-aligned expert specialization, each expert focusing on its designated propagation sub-condition.

\vfill


\begin{thebibliography}{1}
\bibliographystyle{IEEEtran}

\bibitem{busari2018millimeter}
S. A. Busari {\it{et al.}}, "Millimeter-wave massive MIMO communication for future wireless systems: A survey," {\it{IEEE Communications Surveys Tutorials}}, vol. 20, no. 2, pp. 836-869, Q2. 2018.

\bibitem{giordani2019tutorial}
M. Giordani, M. Polese, A. Roy, D. Castor, and M. Zorzi, "A tutorial on beam management for 3GPP NR at mmWave frequencies," {\it{IEEE Communications Surveys Tutorials}}, vol. 21, no. 1, pp. 173-196, Q1. 2019.

\bibitem{xue2024survey}
Q. Xue {\it{et al.}}, "A survey of beam management for mmWave and THz communications towards 6G," {\it{IEEE Communications Surveys Tutorials}}, vol. 26, no. 3, pp. 1520-1559, Q3. 2024.

\bibitem{wang2024beam}
Y. Wang, Z. Wei, and Z. Feng, "Beam training and tracking in mmWave communication: A survey," {\it{China Communications}}, vol. 21, no. 6, pp. 1-22, Jun. 2024.

\bibitem{giordani2018initial}
M. Giordani, M. Polese, A. Roy, D. Castor and M. Zorzi, "Initial access frameworks for 3GPP NR at mmWave frequencies," in {\it{2018 17th Annual Mediterranean Ad Hoc Networking Workshop (Med-Hoc-Net)}}, Capri, Italy, 2018, pp. 1-8.

\bibitem{giordani2019standalone}
M. Giordani, M. Polese, A. Roy, D. Castor and M. Zorzi, "Standalone and Non-Standalone Beam Management for 3GPP NR at mmWaves," {\it{IEEE Communications Magazine}}, vol. 57, no. 4, pp. 123-129, April 2019.

\bibitem{lin2022novel}
K.-H. Lin and K.-H. Liu, "A novel beam alignment scheme for mobile millimeter-wave communications based on compressed sensing aided-Kalman filter," {\it{IEEE Open Journal of the Communications Society}}, vol. 3, pp. 1515-1527, 2022.

\bibitem{jayaprakasam2017robust}
S. Jayaprakasam, X. Ma, J. W. Choi, and S. Kim, "Robust beam-tracking for mmWave mobile communications," {\it{IEEE Communications Letters}}, vol. 21, no. 12, pp. 2654-2657, Dec. 2017.

\bibitem{chen2023mmwave}
L. Chen, S. Zhou, and W. Wang, "MmWave beam tracking with spatial information based on extended Kalman filter," {\it{IEEE Wireless Communications Letters}}, vol. 12, no. 4, pp. 615-619, April 2023.

\bibitem{shen2023compressed}
Q. Shen, A. Hu, and J. He, "Compressed sensing and Kalman filter based channel tracking for mmWave massive MIMO systems," in {\it{International Conference on Wireless Communications and Signal Processing (WCSP)}}, Hangzhou, China, 2023, pp. 1073-1078.

\bibitem{khan2023machine}
M. Q. Khan, A. Gaber, P. Schulz, and G. Fettweis, "Machine learning for millimeter wave and terahertz beam management: A survey and open challenges," {\it{IEEE Access}}, vol. 11, pp. 11880--11902, 2023.

\bibitem{zhang2024tbp}
S. Zhang, Q. Yan, T. Li, L. Xiao, and H. Zeng, "TBP: Temporal beam prediction for mobile millimeter-wave networks," {\it{IEEE Internet of Things Journal}}, vol. 11, no. 14, pp. 24960-24972, Jul. 2024.

\bibitem{liu2025sa}
M. Liu, L. Liang, and W. Guan, "SA-TBP: A speed-adaptive temporal beam prediction framework with AI for mmWave communications," {\it{IEEE Wireless Communications Letters}}, vol. 14, no. 12, pp. 3882-3886, Dec. 2025.

\bibitem{ma2021deep}
K. Ma, D. He, H. Sun, and Z. Wang, "Deep learning assisted mmWave beam prediction with prior low-frequency information," in {\it{ICC 2021 - IEEE International Conference on Communications}}, Montreal, QC, Canada, Jun. 2021, pp. 1-6.

\bibitem{dissanayake2023towards}
D. M. C. Dissanayake, "Towards 6G: Beam prediction using convolutional neural network and artificial neural network," in {\it{2023 7th International Conference on Information Technology (InCIT)}}, Chiang Rai, Thailand, 2023, pp. 392-396.

\bibitem{wang2025deep}
P. Wang, K. Ma, Y. Bai, C. Sun, and Z. Wang, "Deep learning assisted mmWave beam prediction with flexible network architecture," {\it{IEEE Transactions on Wireless Communications}}, vol. 24, no. 11, pp. 9435-9448, Nov. 2025.

\bibitem{wang2023deep}
S. Wang, W. Chen, X. Chen, Y. Zhang, and B. Ai, "Deep learning-based beam pair prediction with finite beam quality information," in {\it{2023 IEEE 23rd International Conference on Communication Technology (ICCT)}}, Wuxi, China, 2023, pp. 588-592.

\bibitem{yang2024hierarchical}
J. Yang and W. Zhu and M. Tao and S. Sun, "Hierarchical Beam Alignment for Millimeter-Wave Communication Systems: A Deep Learning Approach,"  {\it{IEEE Transactions on Wireless Communications}}, vol. 23, no. 4, pp. 3541-3556, April 2024.

\bibitem{Hu2024transfer}
Z. Hu, Y. Li, C.Han, "Transfer learning enabled transformer-based generative adversarial networks for modeling and generating terahertz channels,"  {\it{Communications Engineering}}, 3, 153 (2024).

\bibitem{sheng2025beam}
Y. Sheng {\it{et al.}}, "Beam prediction based on large language models," {\it{IEEE Wireless Communications Letters}}, vol. 14, no. 5, pp. 1406-1410, May 2025.

\bibitem{liu2025large}
W. Liu {\it{et al.}}, "Large-model AI for near-field beam prediction: A CNN-GPT2 framework for 6G XL-MIMO," {\it{IEEE Transactions on Wireless Communications}}, vol. 25, pp. 15149-15165, 2026.

\bibitem{zhang2025multimodal}
K. Zhang {\it{et al.}}, "Multimodal deep learning-empowered beam prediction in future THz ISAC systems," in {\it{2025 IEEE 36th International Symposium on Personal, Indoor and Mobile Radio Communications (PIMRC)}}, Istanbul, Turkiye, 2025, pp. 1-6.

\bibitem{shazeer2017outrageously}
N. Shazeer {\it{et al.}}, "Outrageously Large Neural Networks: The Sparsely-Gated Mixture-of-Experts Layer," in {\it{International Conference on Learning Representations(ICLR)}}, Toulon, France, 2017.


\bibitem{lei2026llm}
J. Lei {\it{et al.}}, "LLM-MM: End-to-end robust multimodal beam prediction for 6G V2X networks via MoE-LoRA adaptation," {\it{IEEE Journal on Selected Areas in Communications}}, vol. 44, pp. 2964-2977, 2026.

\bibitem{liu2025llm4wm}
X. Liu, S. Gao, B. Liu, X. Cheng and L. Yang, "LLM4WM: Adapting LLM for Wireless Multi-Tasking," {\it{IEEE Transactions on Machine Learning in Communications and Networking}}, vol. 3, pp. 835-847, 2025.

\bibitem{zhu2010adaptive}
H. Zhu and J. Wang, "Adaptive Chunk-Based Allocation in Multiuser OFDM Systems," in {\it{2010 IEEE Wireless Communication and Networking Conference}}, Sydney, NSW, Australia, 2010, pp. 1-6.

\bibitem{larsson2014mimo}
E. G. Larsson {\it{et al.}}, "Massive MIMO for next generation wireless systems," {\it{IEEE Communications Magazine}}, vol. 52, no. 2, pp. 186-195, February 2014.

\end{thebibliography}
\end{document}